\documentclass[12pt]{article}
\pdfoutput=1
\usepackage{hyperref}
\usepackage{graphicx}
\usepackage{graphics}
\usepackage{dsfont}
\usepackage{epsfig}
\usepackage{amsmath,amssymb,amsthm,amscd}
\usepackage{simpler-wick}

\newcommand{\Tr}{\textrm{Tr}}

\def\ket{\rangle}
\def\bra{\langle}

\newcommand{\be}{\begin{equation}}
\newcommand{\ee}{\end{equation}}
\newcommand{\ba}{\begin{aligned}}
\newcommand{\ea}{\end{aligned}}

\numberwithin{equation}{section}

\begin{document}
\begin{titlepage}
%{}~ \hfill\vbox{ \hbox{} }\break

\rightline{USTC-ICTS/PCFT-21-23}

\vskip 3 cm

\centerline{\Large 
\bf  
Boson-Fermion Correspondence   } 
\vskip 0.2 cm
\centerline{\Large 
\bf  
 and Holomorphic Anomaly Equation  } 
\vskip 0.2 cm
\centerline{\Large 
\bf 
in 2d Yang-Mills Theory on Torus  }

\vskip 0.5 cm

\renewcommand{\thefootnote}{\fnsymbol{footnote}}
\vskip 30pt \centerline{ {\large \rm 
Min-xin Huang\footnote{minxin@ustc.edu.cn}  
} } \vskip .5cm  \vskip 20pt 

\begin{center}
{Interdisciplinary Center for Theoretical Study,  \\ \vskip 0.1cm  University of Science and Technology of China,  Hefei, Anhui 230026, China} 
 \\ \vskip 0.3 cm
{Peng Huanwu Center for Fundamental Theory,  \\ \vskip 0.1cm  Hefei, Anhui 230026, China} 
\end{center}

\setcounter{footnote}{0}
\renewcommand{\thefootnote}{\arabic{footnote}}
\vskip 40pt
\begin{abstract}

Recently,   Okuyama and Sakai proposed a novel holomorphic anomaly equation for the partition function of 2d Yang-Mills theory on a torus, based on an anholomorphic deformation of the propagator in the bosonic formulation. Using the boson-fermion correspondence, we derive the formula for the deformed partition function in fermionic description and give a proof of the holomorphic anomaly equation.

\end{abstract}

\end{titlepage}
\vfill \eject

%%%%%%%%%%%%%%%%%%%%%%%%%%%%%%%%%%%%%%%%%%%%%%%%%%%%%%%%%%%%%

\newpage

\baselineskip=16pt

\tableofcontents

\section{Introduction}

The 2d Yang-Mills theories are  useful toy models for understanding the physical properties of the more realistic 4d Yang-Mills theories. In particular, they were studied as the early examples of the gauge/string duality before the more famous AdS/CFT correspondence appeared, see e.g. \cite{Witten:1992xu, Gross:1993hu}. We are interested in the $U(N)$ Yang-Mills theory on a torus, whose large $N$ expansion turns out to be equivalent to topological strings on elliptic curves. In this case, the partition function has remarkable modular properties, studied in the early literature, e.g. \cite{Douglas:1993wy, 10.1007/978-1-4612-4264-2_5, 10.1007/978-1-4612-4264-2_6, Rudd:1994ta, Cordes:1994fc}.

The partition function can be computed in two formulations in terms of free fermions and bosons. This is probably one of the simplest examples of mirror symmetry where the fermionic and bosonic formulations roughly correspond to the topological A-model and B-model on the elliptic curve. The higher genus mirror symmetry of the partition function including the gravitational descendants has been mathematically proven by Li  in \cite{Li:2011mx}.  The topological string partition functions on Calabi-Yau three-folds satisfies the powerful Bershadsky-Cecotti-Ooguri-Vafa (BCOV) holomorphic anomaly equations \cite{Bershadsky:1993cx}. However, in the case of elliptic curve with modulus $\tau$, although the partition function can be written in terms of quasi-modular forms, i.e. polynomials of Eisenstein series $E_2, E_4, E_6$, the naive anholomorphic deformation by replacing the holomorphic $E_2$ with the anholomorphic but modular $\hat{E}_2(\tau) \equiv E_2(\tau) -\frac{3}{\pi {\rm Im} (\tau) } $ does not give a nice holomorphic anomaly equation similar to the BCOV equation. See some studies in \cite{Dijkgraaf:1996iy}. This seems very different from most examples in Seiberg-Witten theories and topological strings on elliptic Calabi-Yau three-folds, where one can show that the holomorphic anomaly and modular anomaly equations are essentially equivalent \cite{Huang:2006si, Huang:2013eja, Huang:2020dbh}.

Okuyama and Sakai proposed an anholomorphic deformation of the propagators in the Feynman diagrams of the bosonic formulation, based on some physical arguments \cite{Okuyama:2019rqn}. Although the partition function is not modular with this deformation, a nice BCOV-like holomorphic anomaly equation can be obtained. In this paper, we derive the corresponding formula in fermionic formulation, which was also already conjectured in \cite{Okuyama:2019rqn}. It is then easy to prove the conjectured holomorphic anomaly equation. We will use the well-known boson-fermion correspondence, where some nice expositions are available in e.g. \cite{1791585, Li:2011mx}. A classic introduction to the theory of (quasi)-modular forms is \cite{Zagierbook}.    

The paper is organized as the followings.  In Sec. \ref{bosonsection} we review the formulation of holomorphic anomaly equation in \cite{Okuyama:2019rqn}. In Sec. \ref{fermionsection}, after briefly reviewing the boson-fermion correspondence, we derive the formula for the deformed partition function in the fermionic formulation. It is then straightforward to prove the  holomorphic anomaly equation. in Sec. \ref{conclusion} we conclude with some  potential future directions.

\section{Bosonic Formulation} \label{bosonsection}

The partition function can be expressed by the path integral of a compactified boson field $\varphi$ on the torus with a cubic interaction
\be  \label{partitionZ}
\mathcal{Z} = \int \mathcal{D} \varphi \exp[ \int_{T^2} (\bar{\partial} \varphi \partial \varphi +\frac{g_s}{6} (\partial\varphi)^3 )] .
\ee
We can expand the cubic interaction for small $g_s$ and compute by Feynman diagrams. The propagator is given by the two-point function on torus, which is basally determined by the double periodicity and singularities, in terms of the Weierstrass's elliptic function and Eisenstein series 
\be \label{propagator2.2}
\bra \partial\varphi (z_1) \partial\varphi (z_2) \ket = - \wp(z_1-z_2) -\frac{E_2}{12} +S ,
\ee  
where $S$ is an anholomorphic parameter relating to the imaginary part of the torus modulus 
\be 
S:= \frac{1}{4\pi {\rm Im} (\tau)} . 
\ee
The appearance of $S$ depends on whether one takes into account of the winding mode contributions \cite{Douglas:1993wy, Datta:2014zpa}. The partition function would be entirely holomorphic if one omits the $S$-dependence in the propagator. Here we follow the prescription in \cite{Okuyama:2019rqn} of keeping the $S$-dependence, so that a BCOV-like holomorphic anomaly equation can be obtained. 

The partition function can be expanded in terms of free energy 
\be 
\log(\mathcal{Z}) = \mathcal{F} = \sum_{g=1}^{\infty} g_s^{2g-2} \mathcal{F}_g. 
\ee
This corresponds to the large $N$ expansion of 2d $U(N)$ gauge theory where the genus $g$ appears in the powers of $N$.  The genus one term, which is independent of the cubic coupling constant $g_s$, comes from the normalization of the path integral with the quadratic action. It is convenient to also include the anholomorphic dependence so that it is modular 
\be  \label{genusone} 
\mathcal{F}_1 = \frac{1}{2} \log( S) -\log \eta(\tau),
\ee 
where $\eta(\tau) = Q^{\frac{1}{24}} \prod_{n=1}^{\infty} (1-Q^n)$ with $Q:=e^{2\pi i \tau}$ is the well-known Dedekind  eta function. The shift of the genus one free energy by an additive constant will not affect our calculations here.

The higher genus free energy $\mathcal{F}_g$ is then computed by summing over all connected Feynman diagrams with $2g-2$ cubic vertices and $3g-3$ propagators. The propagator (\ref{propagator2.2}) needs to be integrated over the torus, and for exploring the relation with fermionic formulation, it is convenient to express it as  a power series of $Q$ \cite{bohm2017tropical}. The contribution of a Feynman diagram $\Gamma$ is 
\be
I_\Gamma = \oint \prod_{i=1}^{2g-2} \frac{d x_i} {2\pi i x_i} \prod_{k=1}^{3g-3} G(x^{+}_k/x^{-}_k), 
\ee
where the propagator is given by 
\be \label{propa}
G(x) = \sum_{n=1}^{\infty} \frac{n(x^n +x^{-n} Q^n) }{1-Q^n} +S.  
\ee
Each vertex is associated with a $x_i$ integral. The $x_{\pm}$ in the integral correspond to the two vertices connected by the propagator. Notice the propagator is not invariant under the transformation $x\rightarrow \frac{1}{x}$. A prescription is to assign an order to the $2g-2$ vertices, then for the propagator connecting different vertices, one can always choose $x_{+}$ to be the vertex with the higher order. For the propagator connecting two edges in the same vertex, a zeta-function regularization $\zeta(-1) = -\frac{1}{12}$ is needed to compute the expression 
\be  \label{G1}
G(1) = \sum_{n=1}^{\infty} n +  \sum_{n=1}^{\infty} \frac{2n Q^n}{1-Q^n} +S = -\frac{E_2}{12} +S . 
\ee  
It is familiar in Feynman diagram calculations that the same Feynman rules also apply to the computations of partition functions by simply including the disconnected Feynman diagrams.

As an example, the genus two free energy has contributions from two Feynman diagrams. The result is computed in  \cite{Okuyama:2019rqn} as
\be \label{genustwo}
\mathcal{F}_2 = \frac{5}{24} S^3 - \frac{E_2}{48} S^2 -\frac{E_2^2-2E_4}{1152} S +\frac{5E_2^3 -3E_2E_4 -2E_6} {51840} .   
\ee
One can check that the $E_2$ and $S$ can not be completely combined into the modular $\hat{E}_2$. However, there is a nice holomorphic anomaly equation. One introduces the derivative $D:=Q\partial_Q =\frac{1}{2\pi i }\partial_\tau$, so we have 
\be  \label{drule}
D S = S^2, ~~~ DQ^n =n Q^n.
\ee 
The derivative of genus one free energy is 
\be
D \mathcal{F}_1  =\frac{1}{2} S -\frac{E_2}{24} . 
\ee
The anholomorphic dependences in the free energy and partition function come entirely from the $S$ parameter. It is then easy to check the following genus two holomorphic anomaly equation 
\be 
2\partial_S \mathcal{F}_2  = (D+S) D\mathcal{F}_1 + (D\mathcal{F}_1)^2 . 
\ee
Some higher order free energies were calculated in \cite{Okuyama:2019rqn} and similar holomorphic anomaly equations were checked. It is most convenient to organize the holomorphic anomaly equation in terms of the partition function as 
\be  \label{anomaly} 
(2\partial_S -S^{-1}) \mathcal{Z} = g_s^2 (D+S)D \mathcal{Z}. 
\ee
This looks tantalizingly similar to the celebrated BCOV equation.  

\section{Fermionic Formulation} \label{fermionsection}

\subsection{Boson-Fermion Correspondence and Feynman Rules} 
We briefly review the boson-fermion correspondence, following mostly the notation in \cite{Li:2011mx}. The system of free fermions is described by two sets of Grassmann operators with half integer indices, satisfying the following anti-commutation relations 
\be  \label{anticommute} 
\{b_n , c_m \} =\delta_{m+n,0} ,~~ \{b_n , b_m \} = \{c_n , c_m \}  = 0  ,~~~ m,n \in \mathbb{Z}+\frac{1}{2} . 
\ee
The operators with positive and negative indices are regarded as annihilation and creation operators. The vacuum state $|0 \ket$ is annihilated by all annihilation operators 
\be 
b_n|0\ket = c_n |0\ket =0, ~~~ n>0. 
\ee
One can then construct the linear basis of the Fock space by acting the creation operators on the vacuum. The dual states can be defined with the dual vacuum $\bra 0 |$ annihilated by negative mode operators from the left. The dual state of $b_{-n} |0\ket$ is  $\bra 0 |c_{n} $ for $n>0$ so that the inner product is normalized to be one. The charge of a state is defined to be the difference of the numbers of $b,c$ creation operators acting on the vacuum. The Hilbert space is decomposed into the charged sectors as $H = \oplus_{p\in \mathbb{Z}} H_p$.  We will be interested in the zero-charge sector $H_0$, which are constructed by acting the same number of $b$ and $c$ creation operators on the vacuum.  

It is convenient to define the fermionic fields which are generating functions of the fermionic operators 
\be 
b(z) = \sum_{n\in \mathbb{Z}+\frac{1}{2}} b_n z^{-n-\frac{1}{2}},~~~ c(z) = \sum_{n\in \mathbb{Z}+\frac{1}{2}} c_n z^{-n-\frac{1}{2}}. 
\ee
The duality of states can be conveniently denoted as $b(z)^{\dagger} = \bar{z}^{-1} c(1/\bar{z})$. We have the following relation, appeared in the well-known operator product expansion (OPE) in conformal field theory 
\be  \label{fermionOPE}
b(z) c(w) = \frac{1}{z-w} + : b(z) c(w) :, ~~~~ {\rm for} ~|z|>|w|,
\ee
where $::$ is the normal ordering which moves the annihilation operator to right, with sign for the fermionic case.

The system of free bosons can be constructed from the above fermionic system by 
\be  \label{alphafield}
\alpha (z) = \sum_{n\in \mathbb{Z}} \alpha_n z^{-n-1} = :b(z) c(z): . 
\ee
The bosonic mode operators can be expressed in terms of fermions 
\be \label{alphamode}
\alpha_n = \sum_{k\in \mathbb{Z}+\frac{1}{2}} : b_k c_{n-k}: . 
\ee
They form a set of infinite harmonic oscillators with the commutation relations 
\be
[\alpha_m, \alpha_n]=m \delta_{m+n, 0 },~~~ m,n\in \mathbb{Z}.
\ee
Again we identify the bosonic operators with positive and negative indices as annihilation and creation operators, while for the zero mode operator $\alpha_0$, it is easy to see that a  fermionic state with charge $p$ is an eigenstate of  with eigenvalue $p$. For each sector $H_p$, there is a unique fermionic state which is created by the lowest fermionic creation operators and is annihilated by all bosonic annihilation operators. We can then construct the bosonic Fock space by acting the creation operators on the vacuum state in each sector $H_p$. In this way there is an isomorphism of the bosonic and fermionic Hilbert spaces.

One defines another bosonic field $\phi(z)$ by 
\be 
\phi(z) = P + \alpha_0 \log z + \sum_{n\neq 0} \frac{\alpha_{-n}} {n} z^n ,
\ee
which is related to the bosonic field in (\ref{alphafield}) by $\alpha(z) =\partial_z \phi(z)$.  The momentum operator $P$ is the conjugate of the zero mode operator $[\alpha_0, P]=1$. Its exponential $e^P$ increases the charge of a state by one. Since we will be interested only in the zero charge sector, the operators $P, \alpha_0$ will not be important in our context. It is straightforward to check the relation, which is familiar from OPE 
\be 
\phi(z) \phi(w)  =\log(z-w) + : \phi(z) \phi(w) : , ~~~~ {\rm for} ~|z|>|w|,
\ee
where $::$ is the bosonic normal ordering. 

The theorem of boson-fermion correspondence \cite{1791585} states that the fermionic fields can be expressed 
\be 
b(z) = :e^{\phi(z)} :, ~~~ c(z) = :e^{-\phi(z)} :,
\ee
It is also familiar in CFT computations that 
\be 
: e^{\phi(z)}: :e^{-\phi(w)}: =\frac{1}{z-w} :e^{\phi(z) -\phi(w)}: . 
\ee  
Together with the fermionic formula (\ref{fermionOPE}), we arrive at the following useful formula 
\be \ba \label{useful3.14} 
\mathcal{E}(z, \lambda) &\equiv \frac{1}{(e^{\lambda/2} - e^{-\lambda/2})z}  :e^{\phi(e^{\lambda/2} z) -\phi(e^{-\lambda/2}z)}:  \\ &
= :b(e^{\lambda/2 } z) c(e^{-\lambda/2 } z) : +\frac{1}{(e^{\lambda/2} - e^{-\lambda/2})z} .
\ea \ee
The exponent in bosonic expression can be written in terms of mode expansion 
\be 
\phi(e^{\lambda/2} z) -\phi(e^{-\lambda/2}z) = \lambda \alpha_0 + \sum_{n\neq 0} \frac{\alpha_{-n} } {n} z^n (e^{n\lambda/2} - e^{-n\lambda/2}). 
\ee 
One can then expand the above formula (\ref{useful3.14}) as Laurent series of $\lambda, z$ and compare the coefficients. In this way, many non-trivial identities between the bosonic and fermionic mode operators can be obtained. We will focus only on the $z^{-1}$ coefficient, which does not change the energy of a state. The formula for fermionic modes is 
\be 
\oint \frac{dz}{2\pi i }  \mathcal{E}(z, \lambda) = \sum_{k\in \mathbb{Z}+\frac{1}{2}} e^{\lambda k} :b_{-k} c_k : + \frac{1}{e^{\lambda/2} - e^{-\lambda/2}} . 
\ee 

Comparing the coefficients of $\lambda z^{-1} $ in the formula (\ref{useful3.14}) gives the the zero mode Virasoro operator, or the energy  
\be  \label{L0formula}
L_0 = \frac{1}{2} \alpha_0^2 +\sum_{n=1}^{\infty} \alpha_{-n} \alpha_n = \sum_{k\in \mathbb{Z}^{\geq 0} + 1/2 } k(b_{-k}  c_k +c_{-k} b_k ) .
\ee  
This familiar relation can be also directly derived using (\ref{alphamode}) after a careful calculation with the anti-commutation relations. The generating function with trace over the zero charge sector is related to the Dedekind eta function 
\be 
\Tr_{H_0} Q^{L_0- \frac{1}{24}} = \eta(\tau)^{-1}   ,
\ee 
which can be computed by both bosonic and fermionic formulas in (\ref{L0formula}), whose equivalence is implied by the classic Jacobi triple identity. The other Virasoro operators can be also defined in terms of both bosons and fermions but will not be needed in our context.

For our purpose, we will also compare the coefficients of $\lambda^2 z^{-1} $.  In this case the bosonic exponential function in (\ref{useful3.14})  needs to be expanded to the cubic  order. Since we will only consider the zero charge sector, we can omit the $\alpha_0$ in the formulas. We denote the operator 
\be \ba  \label{O}
\mathcal{O} & \equiv  \frac{1}{6}  \oint \frac{dz}{2\pi i z} :(  \sum_{n\neq 0 }  \alpha_{-n} z^n)^3 :   \\ &
=  \sum_{k\in \mathbb{Z} +\frac{1}{2} } \frac{k^2}{2} :b_{-k} c_k:  . 
\ea \ee
The trace of $Q^{L_0-\frac{1}{24}} \mathcal{O}^{2g-2}$ over the zero charge sector $H_0$ gives the $g_s^{2g-2}$ order contributions to the partition function (\ref{partitionZ}) without the $S$ deformation. This can be computed according to the Wick theorem by contracting the bosonic mode operators, including the disconnected diagrams. Usually the Wick contraction for a bosonic harmonic oscillator is between a creation and annihilation operator with the creation operator on the right. We are using an ``effective Wick contraction"  with trace, where the formula for two modes is computed as 
\be \ba \label{Wick3.18}
 \Tr _{H_0} (Q^{L_0-\frac{1}{24}}  \alpha_m \alpha_n ) &= 
 \delta_{m+n,0} \eta(\tau)^{-1}  (1-Q^n)  \sum_{k=0}^{\infty} nk  Q^{nk}  \\
& =  \delta_{m+n,0} \eta(\tau)^{-1}  \frac{ n Q^n}{1-Q^n}  ,  & \text{for } n\geq 0 . 
\ea \ee
While the case of $n<0$ can be obtained by the canonical commutation relation with an extra term 
\be \ba
 \Tr _{H_0} (Q^{L_0-\frac{1}{24}}  \alpha_m \alpha_n ) &= 
 \delta_{m+n,0} \eta(\tau)^{-1}  (1-Q^n)  \sum_{k=0}^{\infty} n(k+1)  Q^{nk}  \\
& =  \delta_{m+n,0} \eta(\tau)^{-1}  \frac{m}{1-Q^m}  ,  & \text{for } n< 0 . 
\ea \ee
It is then straightforward to compute 
\be \ba \label{compute3.17}
&~~ \Tr _{H_0} (Q^{L_0-\frac{1}{24}} \sum_{m,n} \alpha_m \alpha_n x_{-}^{-m} x_{+}^{-n})   =
 \eta(\tau)^{-1}  \sum_{n=1}^{\infty} \frac{n [    (\frac{x_{+}}{x_-}) ^{n}   +(\frac{x_{-}}{x_+}) ^{n} Q^n ]}{1-Q^n} , 
\ea \ee
which matches with the propagator (\ref{propa}) without the $S$-deformation, while the factor $ \eta(\tau)^{-1}$ accounts for the contribution of genus one free energy (\ref{genusone}). The effective Wick contraction between two edges from the same vertex does not contribute in this case since we restrict to the zero charge sector.  

The calculations (\ref{compute3.17}) can be independently done for all possible effective Wick contractions of the edges of the cubic vertices in (\ref{O}), and there is only one overall factor of $ \eta(\tau)^{-1}$ for a Feynman diagram.  To illustrate, let us for simplicity consider the case of $\alpha_{-1}, \alpha_1$ operators.   Focusing on the sector of Hilbert space created by $\alpha_{-1}$, we denote $|k\ket = \frac{(\alpha_{-1})^k}{\sqrt{k!}}|0\ket$ the properly normalized $k$-particle state. Then we compute schematically  
\be \ba
\sum_{k=0}^{\infty}  \langle
 \wick{
         k |  Q^{\alpha_{-1} \alpha_1}   \cdots \c1 \alpha_{-1}\cdots  \c1 \alpha_{1} \cdots |   k }
\rangle =  \sum_{k=0}^{\infty} k 
\bra k |  Q^{\alpha_{-1} \alpha_1 }   \cdots  |   k  \ket =   Q\partial_Q \sum_{k=0}^{\infty}  
\bra k |  Q^{\alpha_{-1} \alpha_1 }   \cdots  |   k  \ket. 
\ea \ee
The calculation can be understood as first performing an implicit ``usual Wick contraction" of $\alpha_1$ with the creation operators in $|k\ket$, then we use the $\alpha_{-1}$ to fill in for the creation operator. So it looks effectively we are contracting $\alpha_{-1}, \alpha_{1}$ as displayed in the above expression. After all contractions are done, the trace is simply $ \sum_{k=0}^{\infty}  \bra k |  Q^{\alpha_{-1} \alpha_1 }    |   k  \ket    =(1-Q)^{-1}$. So the effect of the $\alpha_{-1}, \alpha_1$ contraction can be derived as $ (1-Q) Q\partial_Q (1-Q)^{-1} = \frac{Q}{ 1-Q}$, consistent with (\ref{Wick3.18}) for $n=1$, independently of the other possible operators denoted as  $\cdots$ in the above expression. Similarly, the effective Wick contraction of $\alpha_{1}, \alpha_{-1}$ with $\alpha_{-1}$ on the right gives a factor of $ \frac{1}{ 1-Q}$. 

As a check of the somewhat tricky reasonings, we can compute directly a simple  example 
\be
\sum_{k=0}^{\infty}   \bra k |  Q^{\alpha_{-1} \alpha_1 } (\alpha_{-1} \alpha_1 )^2    |   k  \ket   = \sum_{k=0}^{\infty}  k^2 Q^k = \frac{Q(1+Q)}{(1-Q)^3}. 
\ee
On the other hand, we can compute by the effective Wick contractions as
\be \ba 
&\sum_{k=0}^{\infty}   \bra k |  Q^{\alpha_{-1} \alpha_1 } (  \wick{ \c1 \alpha_{-1} \c1 \alpha_1 } \wick { \c1 \alpha_{-1} \c1 \alpha_1}  +  \wick{ \c1 \alpha_{-1} \c2 \alpha_1   \c2 \alpha_{-1} \c1 \alpha_1    } |k  \ket   \\ &= (1-Q)^{-1} [ \frac{Q^2 }{(1-Q)^2} +  \frac{Q }{(1-Q)^2} ] \\ &=\frac{Q(1+Q)}{(1-Q)^3},
\ea \ee
which gives the same result. 

So the effective Wick contractions give the same Feynman rules described in the previous Sec. \ref{bosonsection} without the $S$-deformation.  The paper \cite{Li:2011mx} considers  the much more general cases with insertions of descendant operators in the correlation functions in the mathematical definition of Gromov-Witten invariants of elliptic curves. Here we focus on the special case where the first Chern class of the line bundles of marked points in the Riemann surface have degree one. In this case the genus $g$ contribution must have $2g-2$ insertions, enforced by the selection rule due to the dimension of the virtual fundamental class of the moduli space. This corresponds to the $2g-2$ cubic vertices in the Feynman diagrams.

The computations of the trace of $Q^{L_0-\frac{1}{24}} \mathcal{O}^{2g-2}$ with the fermionic operators in the formula (\ref{O}) give the free fermion formulation of the partition function (without the $S$-deformation), well studied in e.g.  \cite{Douglas:1993wy, 10.1007/978-1-4612-4264-2_6}. We will give the detailed analysis for the case with $S$-deformation in the next subsection \ref{secdeformed}.

\subsection{Deformed Partition Function} \label{secdeformed}

%To produce the $S$-dependence in the propagator (\ref{propa}), we introduce another harmonic oscillator $[a,a^{\dagger}]=1$, which is completely independent of the boson-fermion systems discussed above. It is easy to compute the vacuum expectation value 
%\be 
%\bra 0 | (a+a^\dagger)^{2n-1} |0 \ket =0, ~~ \bra 0 | (a+a^\dagger)^{2n} |0 \ket =(2n-1)!!, ~~~ n\in \mathbb{N}. 
%\ee
%The formula can be understood as the number of Wick contractions between $a+a^\dagger$, and can be also computed using the ground state wave function of the harmonic oscillator.  

To produce the $S$-dependence in the propagator (\ref{propa}), we introduce an auxiliary parameter $a$. It is easy to compute the Gaussian integrals
\be  \label{gauss}
\int_{-\infty}^{\infty} \frac{e^{-\frac{a^2}{2}} da}{\sqrt{2\pi}}  a^{2n}=(2n-1)!!, ~~~ n\in \mathbb{N}.  
\ee 
The formula can be understood in terms of Feynman rule as the number of Wick contractions between $a$'s. 

We multiply the formula (\ref{useful3.14}) by a factor $e^{\lambda a\sqrt{S} }$, and again compare the coefficients of $\lambda^2 z^{-1}$ in the Laurent expansion. We denote the operator as  
\be \ba  \label{Os}
\mathcal{O}(S) & \equiv  -\frac{1}{24} aS^{\frac{1}{2}}  +\frac{1}{6} \oint \frac{dz}{2\pi i z} :( aS^{\frac{1}{2}} + \sum_{n\neq 0 }  \alpha_{-n} z^n)^3 : \\ &
=  \sum_{k\in \mathbb{Z} +\frac{1}{2} } \frac{(k+a\sqrt{S})^2}{2} :b_{-k} c_k:  
+ \frac{1}{6} a^3 S^{\frac{3}{2}} - \frac{1}{24} aS^{\frac{1}{2}} . 
\ea \ee
Performing the integral $\int_{-\infty}^{\infty} \frac{e^{-\frac{a^2}{2}} da}{\sqrt{2\pi}}  $ on the auxiliary parameter $a$ will enforce a Wick contraction between $S^{\frac{1}{2}}$'s in the cubic vertex, giving rise the deformed propagator (\ref{propa}) with the $S$ term.  Our prescription has an extra term $-\frac{1}{24} aS^{\frac{1}{2}} $ besides the cubic vertex. It accounts for the normal ordering constant when two edges from the same vertex are Wick contracted.  In this case, we can calculate 
\be \ba
&~~ \Tr _{H_0} (Q^{L_0-\frac{1}{24}} \sum_{m,n } :\alpha_{m} \alpha_{n}: z^{-m-n} ) 
 = \eta(\tau)^{-1}  \sum_{n=1}^{\infty} \frac{2n Q^n }{1-Q^n} .
\ea \ee
Comparing with (\ref{compute3.17}) for $x_{+}=x_{-}$ or (\ref{G1}), there is a difference with normal ordering constant $\sum_{n=1}^{\infty}  n =-\frac{1}{12}$. This is exactly compensated by the extra term $-\frac{1}{24} aS^{\frac{1}{2}} $, since $3\times \frac{1}{6} (-\frac{1}{12}) = -\frac{1}{24}$. 

Now using the fermionic formula in (\ref{Os}), we can finally write the deformed partition function (\ref{partitionZ}) in terms of summation over 2d Young tableaux
\be \ba \label{compute3.21}
\mathcal{Z} &= \sqrt{S} \int_{-\infty}^{\infty} \frac{e^{-\frac{a^2}{2}} da}{\sqrt{2\pi}}  \sum_{g=1}^{\infty} \frac{g_s^{2g-2} }{(2g-2)!}
 \Tr_{H_0}  [Q^{L_0-\frac{1}{24}} \mathcal{O}(S)^{2g-2}] \\
 &= \sqrt{S}  \int_{-\infty}^{\infty} \frac{e^{-\frac{a^2}{2}} da}{\sqrt{2\pi}} \oint \frac{dx}{2\pi i x}
Q^{-\frac{1}{24}} e^{g_s (\frac{1}{6} a^3 S^{\frac{3}{2}} - \frac{1}{24} aS^{\frac{1}{2}} ) }    \\
 & \times   \prod_{p\in\mathbb{Z}^{\geq 0} +\frac{1}{2}} 
 [1+ x Q^p e^{g_s(\frac{p^2}{2} + p a \sqrt{S} )} ]  [1+ x^{-1} Q^p e^{g_s(- \frac{p^2}{2} + p a \sqrt{S} )} ]   \\
 &=  \sqrt{S}  \int_{-\infty}^{\infty} \frac{e^{-\frac{a^2}{2}} da}{\sqrt{2\pi}} \sum_{R} 
 Q^{|R|-\frac{1}{24}}  e^{g_s \sum_i R_i(R_i -2i +1) }  e^{g_s [\frac{1}{6} a^3 S^{\frac{3}{2}} +(|R|- \frac{1}{24} )aS^{\frac{1}{2}} ] }  . 
\ea \ee
Some explanations are in orders. We have include a factor of $\sqrt{S}$ to match the convention of genus one free energy in (\ref{genusone}). The last expression sums over all 2d Young tableaux $R$, labelled by a sequence of non-increasing positive integers $R_1\geq R_2 \geq \cdots $. The number of boxes is denoted $|R|= \sum_i R_i $. It is well known that a 2d Young tableau has a Frobenius representation in terms of two sets of strictly decreasing positive half integers as $(p_1, p_2, \cdots, p_n| \tilde{p}_1, \tilde{p}_2, \cdots , \tilde{p}_n)$.  The map can be obtained by dissecting the  Young tableau through its diagonal elements. It is obvious that $|R|= \sum_{i=1}^n (p_i +\tilde{p}_i )$.  There is an isomorphism with the zero charge Hilbert space $H_0$, where the rows of the Young tableau corresponds to the bosonic creation operators, while the Frobenius representation gives the fermionic Fock space. For example, the Frobenius representation $(p_1, p_2, \cdots, p_n| \tilde{p}_1, \tilde{p}_2, \cdots , \tilde{p}_n)$ corresponds to the state   
\be \label{psi}
 |\psi\ket = \prod_{i=1}^n b_{-p_i}  c_{-\tilde{p}_i} |0\ket  .
 \ee

In the second equality in  (\ref{compute3.21}) we have used the fact that the fermionic state corresponding to the Frobenius representation is an eigenstate of the fermionic operator in (\ref{Os}). In fact, for the state (\ref{psi}) it is easy to see 
\be \ba
  \sum_{k\in \mathbb{Z} +\frac{1}{2} } f_k :b_{-k} c_k:  |\psi\ket  =  \sum_{j=1}^n  ( f_{p_j}  -   f_{-\tilde{p}_j}   ) |\psi\ket ,
  \ea \ee
for any  coefficients $f_k$'s.  

The computation in (\ref{compute3.21}) also uses the following formula of the Frobenius representation. For a 2d Young tableau $R$, we put the numbers $0,1, \cdots, R_1-1$ in the first row, then $-1,0, 1, \cdots, R_2-2$ in the second row, etc. The diagonal elements in the Frobenius partition are always 0. Suppose the Frobenius representation is $(p_1, p_2, \cdots, p_n| \tilde{p}_1, \tilde{p}_2, \cdots , \tilde{p}_n)$, we can sum over the numbers in the boxes in two ways to easily prove 
\begin{equation} 
\sum_i p_i^2 - \sum_i \tilde{p}_i^2 =\sum_i R_i (R_i -2i+1) .
\end{equation} 

Since the Frobenius representation of the transposition of $R$ is simply obtained by switching the two sets of half integers, the half integer powers of $S$ always cancel between a Young tableau and its transposition in the formula (\ref{compute3.21}). So it is clear  that the free energies are polynomials of $S$. It is also obvious that the odd $g_s$ power terms vanish by a trivial change of the integration variable $a\rightarrow -a$. 

We can perform some simple checks of the formula (\ref{compute3.21}). For example, we can expand to $g_s^2$ and compare with the  $S^2, S^3$ terms in the genus two formula (\ref{genustwo}).  We find consistent results 
\be \ba 
\mathcal{Z} 
 &=  \sqrt{S}  \int_{-\infty}^{\infty} \frac{e^{-\frac{a^2}{2}} da}{\sqrt{2\pi}} \sum_{R} 
 Q^{|R|-\frac{1}{24}} \{1+ g_s^2 [ \frac{a^6S^3}{72}  +(|R|- \frac{1}{24}) \frac{a^4S^2}{6} +\cdots  ] \}+\mathcal{O}(g_s^4)  \\
 &=  \sqrt{S}  \eta(\tau)^{-1} [1+ g_s^2 ( \frac{5}{24} S^3 - \frac{E_2 }{48} S^2 +\cdots )  +\mathcal{O}(g_s^4) ] . 
\ea \ee

We note that the fermionic formula (\ref{compute3.21}) was already essentially also conjectured by Okuyama and Sakai \cite{Okuyama:2019rqn} in a somewhat different notation. Here we have given a proof using the boson-fermion correspondence formula (\ref{Os}). 

\subsection{Holomorphic Anomaly Equation} 

As in \cite{Okuyama:2019rqn}, it is now straightforward to analyze the holomorphic anomaly equation (\ref{anomaly}) with the fermionic formula (\ref{compute3.21}). Here we perform a more direct computation using the differential rules (\ref{drule} 
\be  \ba \label{formula3.24}
& ~~ \left[ -2\partial_S +S^{-1} + g_s^2 (D+S)D \right]  \mathcal{Z}  \\
& =  \sum_{R} 
  (|R|-\frac{1}{24})^{-\frac{1}{2}}  Q^{|R|-\frac{1}{24} } e^{g_s \sum_i R_i(R_i -2i +1) }  
 \\ &\times \int_{-\infty}^{\infty} \frac{e^{-\frac{a^2}{2}} da}{\sqrt{2\pi}}  e^{\tilde{g}_s (\frac{1}{6} a^3 \tilde{S}^{\frac{3}{2}} + a\tilde{S}^{\frac{1}{2}} ) }   \{ -\tilde{g}_s a(1 +\frac{1}{2}\tilde{S}a^2 ) +\tilde{g}_s^2 \tilde{S}^{\frac{1}{2}} (1  + 2  \tilde{S} +\frac{5}{4}\tilde{S}^2) 
\\ & + \tilde{g}_s^3 \tilde{S}^2 [ a +  \tilde{S} (\frac{7}{4} a+\frac{1}{2}a^3 ) +\frac{9}{8} \tilde{S}^2 a^3 ] +\frac{1}{16} \tilde{g}_s^4 \tilde{S}^{\frac{7}{2}} a^2 (2   +
 \tilde{S}  a^2)^2 \}, 
\ea \ee
where for convenience we have changed variables  
\be
S= \tilde{S}  (|R|-\frac{1}{24}) , ~~~g_s = \tilde{g}_s  (|R|-\frac{1}{24})^{-\frac{3}{2}} .
\ee 
We expect that the integrand in the last two lines in the formula (\ref{formula3.24}) can be written as a total derivative, so that the integral always vanishes. It is not difficult to make an ansatz and find the solution 
\be
\partial_a  \{  \frac{\tilde{g}_s}{4} e^{-\frac{a^2}{2}+ \tilde{g}_s (\frac{1}{6} a^3 \tilde{S}^{\frac{3}{2}} + a\tilde{S}^{\frac{1}{2}} )  } [ 4(1 +  \tilde{S} )+ \tilde{g}_s  \tilde{S} ^{\frac{3}{2}} (4 + 5  \tilde{S} )a+    \tilde{S}  (2 + \tilde{g}_s^2  \tilde{S}^2)a^2 + \tilde{g}_s  \tilde{S} ^{\frac{5}{2}} a^3+ \frac{1}{2} \tilde{g}_s^2  \tilde{S} ^4  a^4] \}. 
\ee 
Of course there seem to be some potential divergences in the integral in (\ref{formula3.24}) as $a\rightarrow \pm \infty$ due to the cubic term $a^3$ in the exponential. However it is at least completely well defined as a formal power series of $g_s$, as mentioned in \cite{Okuyama:2019rqn}, and the fact that the integrand is a total derivative is sufficient to ensure the vanishing of the formal power series.

\section{Conclusion}  \label{conclusion} 

We have derived the fermionic formula (\ref{compute3.21}) for the deformed partition function of 2d Yang-Mills theory on a torus, and prove the conjectured holomorphic anomaly equation (\ref{anomaly}).  It may be interesting to consider further possible generalizations of the holomorphic anomaly equation, e.g. including the gravitational descendants \cite{Li:2011mx}, with $q$-deformation \cite{Aganagic:2004js}. It would be also interesting to explore whether the techniques used here have some interesting applications in more sophisticated situations, e.g. for topological strings on Calabi-Yau three-folds.

As we mentioned, although it is justified by some physical arguments, Okuyama and Sakai's anholomorphic deformation  \cite{Okuyama:2019rqn} does not form a modular combination with the quasi-modular forms. On the other hand, it appears that in the proposed mathematical definitions of BCOV-like theories associated with elliptic curves,  e.g. \cite{Li:2011mx, Costello:2012cy, Li:2020ljm}, the partition functions are always modular. It would be interesting to explore whether there are more useful anholomorphic deformations in more general settings.

\vspace{0.2in} {\leftline {\bf Acknowledgments}}
\nopagebreak

We thank Bao-ning Du, Jun-Hao Li, Gao-fu Ren, Pei-xuan Zeng for helpful discussions, and Sheldon Katz, Albrech Klemm, Yuji Sugimoto, Xin Wang for collaborations on related topics. This work was supported in parts by the national Natural Science Foundation of China (Grants  No.11947301 and No.12047502).

\appendix

\addcontentsline{toc}{section}{References}

%\bibliographystyle{utphys} 
%\bibliography{Reference2DYM}

\providecommand{\href}[2]{#2}\begingroup\raggedright\endgroup

\end{document}